\documentstyle[multicol,prl,epsfig,aps,floats]{revtex}

\begin{document}
\wideabs{
\title{Cholesteric Phase in Virus Suspensions}
\author{Zvonimir Dogic and Seth Fraden}
\address{The Complex Fluids Group, Martin Fisher School of Physics, Brandeis University,
Waltham MA 02454}
 \date{\today}
\maketitle

\begin{abstract} 
We report measurements of the cholesteric pitch and twist elastic
constant $(K_{22})$ in monodisperse suspensions of the rod-like
virus filamentous bacteriophage {\it fd}. Measurements were taken
for concentrations spanning the entire cholesteric region at
several ionic strengths and temperatures. In the limit of high
ionic strength the cholesteric pitch $(P_0)$ scales with
concentration (c) as $P_0 \propto c^{-1.66}$. As the ionic strength
decreases the scaling exponent systematically changes to lower
values.
\end{abstract}}

\newpage
\section{Introduction}

The system with the simplest intermolecular interaction known to
exhibit all the essential features of the nematic state is that of
a hard rod suspension \cite{Onsager49,Vroege92}. Because of its
inherent simplicity, much effort has been put into understanding
the relationship between the microscopic parameters of hard rods
and the resulting liquid crystalline behavior at the macroscopic
level. In nature it often happens that a symmetry of the nematic
phase can be reduced to form a cholesteric phase, where the nematic
director follows a helical path in space. Formation of such a phase
at the macroscopic level is usually associated with chirality of
molecules at the molecular level. Details of how a simple change of
a few atomic positions, required to make a molecule chiral, causes
a drastic change in self organization at the macroscopic level
remains unknown. However, in the continuum limit, where details of
the microscopic interactions are ignored, formation of the
cholesteric phase is  understood as a competition between two
elastic energies. On one hand, the free energy of a chiral nematic
is lowered in a twisted state because of the torque a chiral
molecule exerts on it's neighbor. Such a contribution to the free
energy is characterized by the ``twist'' constant $K_{t}$. On the
other hand, creation of an elastically distorted state
characterized by the usual twist elastic constant $K_{22}$ raises
the free energy \cite{deGennes93}. It follows that the wavelength
of the cholesteric pitch is proportional to the ratio of
$K_{t}/K_{22}$. At present, the challenge lies in calculating the
value of the ``twist'' constant $K_{t}$ for a given molecule with
known microscopic interactions.

Inspired by work of Onsager, Straley made the first attempt to
explain the microscopic origin of the cholesteric phase
\cite{Straley73b,Straley76}. He considered rods with threads of
definite handiness, and for the first time obtained an expression
for $K_{t}$ as a function of the microscopic parameters of a
threaded rod. As in the case of Onsager's hard rods, Straley's
cholesteric phase is purely entropy driven. The non-zero value of
the chiral ``twist'' constant $K_{t}$ is associated with the free
volume gained as threaded rods approach each other at an well
defined angle. This work was later extended to account for
flexibility of rods \cite{Odijk87,Pelcovits96}.

It is not clear if there are lyotropic liquid crystals where the
cholesteric phase is purely entropy driven and therefore most of
the predictions of the Straley model remain untested. An
alternative proposal for the origin of a non-zero $K_{t}$ constant
involves chiral attractions of van der Waals origin
\cite{Issaenko99}. It is likely that for almost all experimental
systems both entropic temperature-independent interactions and
attractive temperature-dependent interactions contribute to the
cholesteric twist, further complicating the problem. Harris and
coworkers noticed that if the threaded rod is allowed to rotate
freely around it's long axis, chirality will be effectively
averaged away and proposed that short-ranged bi-axial correlations
are critical for formation of a cholesteric phase
\cite{Harris97,Harris99}. The implication of their work is that all
mean-field theories, like the one of Straley, do not capture the
essence of cholesteric phase since they ignore all correlations.

\begin{figure}
\centerline{\mbox{\epsfig{file=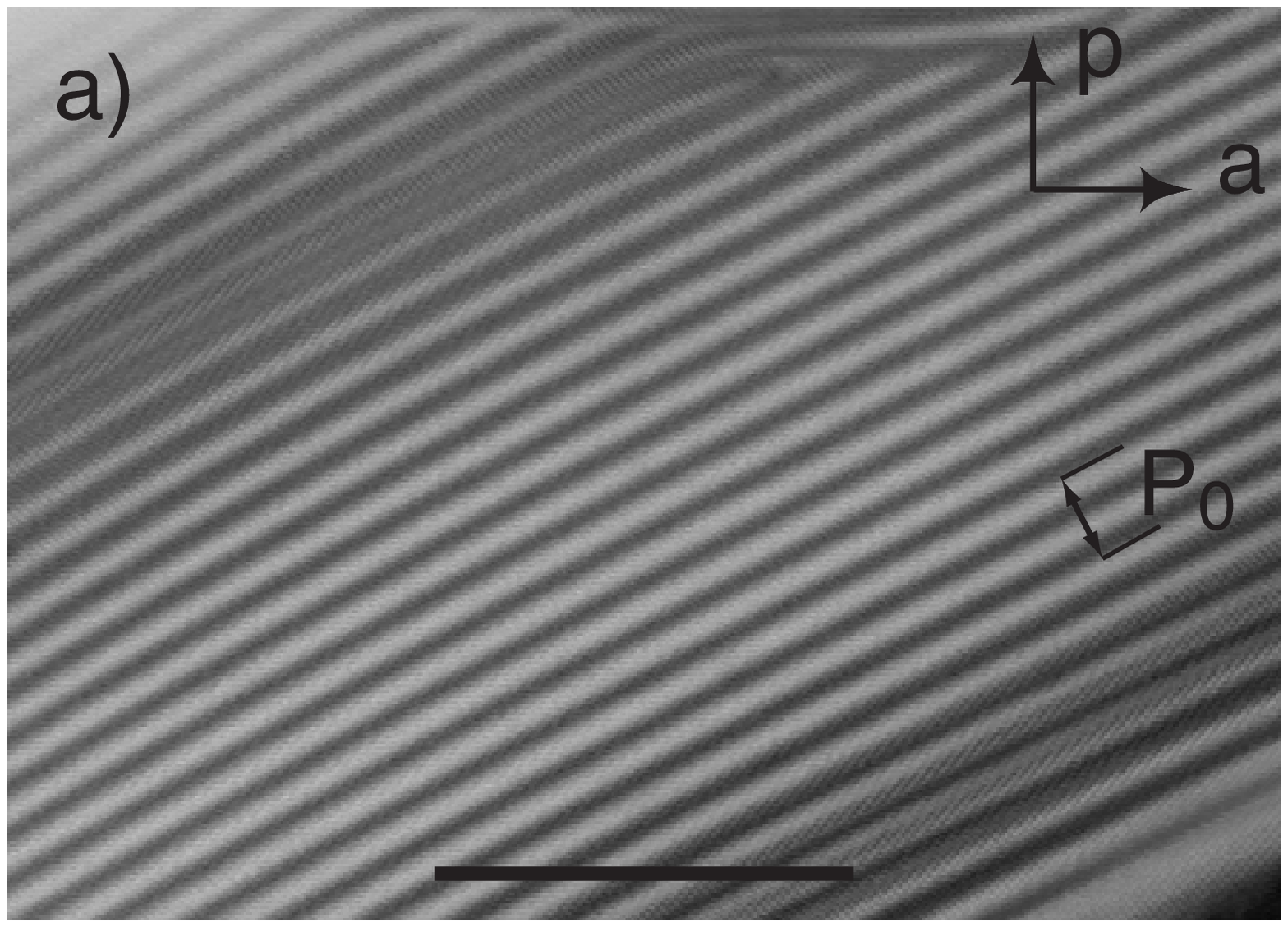,width=8.25cm}}}
\centerline{\mbox{\epsfig{file=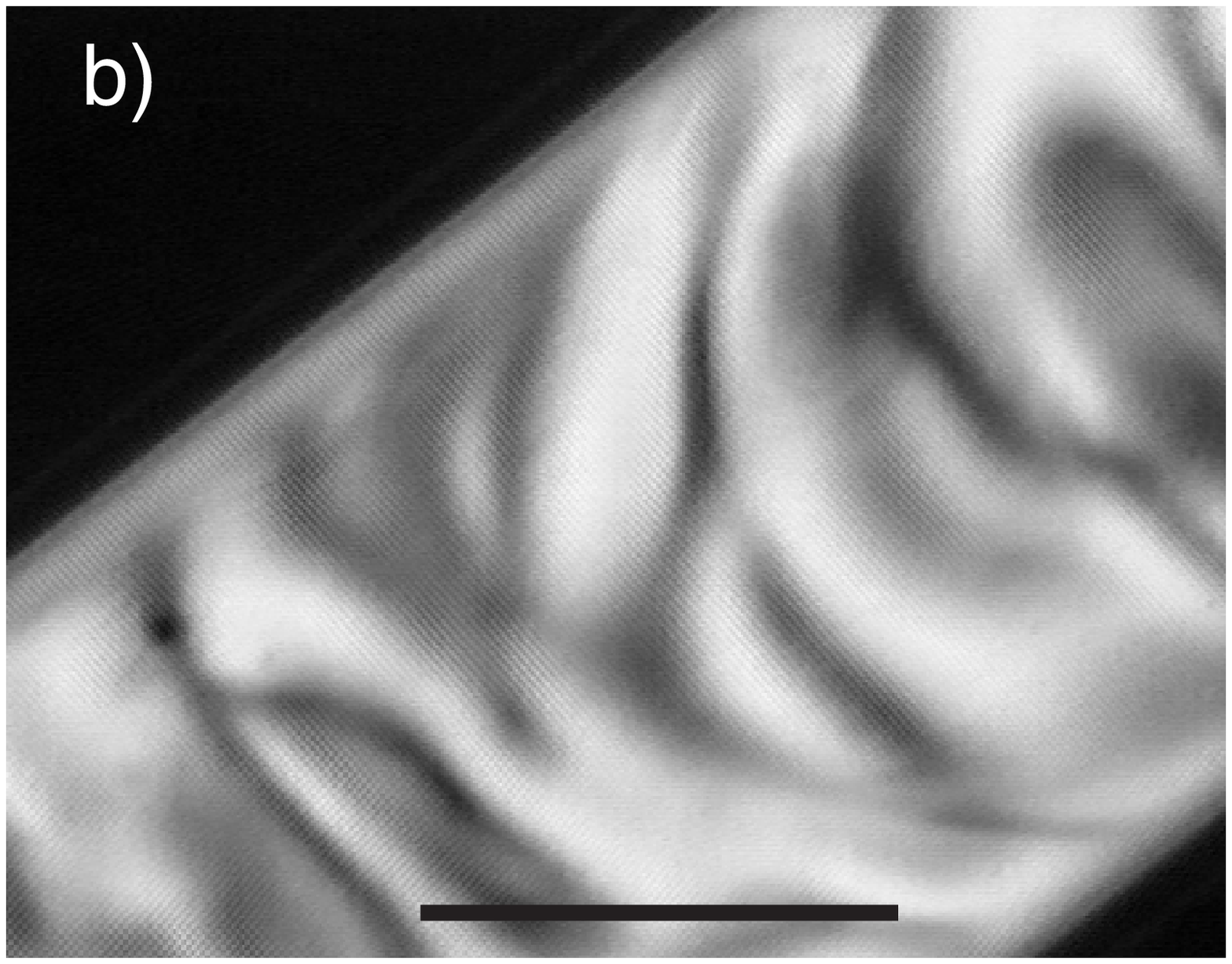,width=8.25cm}}}
\caption{\label{cholesteric} (a) Typical texture of a cholesteric
phase of a liquid crystalline sample of {\it fd} observed with
polarization microscopy. Dark lines correspond to regions where the
rods are perpendicular to the plane of the paper and bright lines
correspond to regions where rods are in the plane of paper. The
length of the cholesteric pitch $(P_0)$ spans two bright and two
dark lines as indicated in the micrograph. The concentration of the
{\it fd} virus is 48 mg/ml and the ionic strength is 8 mM. The
positions of the polarizer (p) and analyzer (a) are indicated in
the upper right corner. (b) Texture observed with a polarization
microscope for a nematic phase formed by filamentous virus Pf1. A
lower limit of the pitch of Pf1 is the capillary diameter of 0.3
mm. Although at the molecular level the two viruses have remarkable
similarity there is no evidence of a cholesteric phase in Pf1 as
indicated by lack of a fingerprint texture. The sample
concentration is 38 mg/ml and the ionic strength is 8 mM.}
\end{figure}

Bacteriophage {\it fd} is a monodisperse rod-like colloid that
forms a cholesteric phase with a characteristic ``fingerprint''
texture shown in Fig.~\ref{cholesteric}a \cite{Lapointe73}.
However, {\it Pf1} a virus with a structure extremely similar to
{\it fd} \cite{Caspar81,Marvin98} does not show any evidence of
forming a cholesteric liquid crystal as shown by the absence of a
``fingerprint'' texture in Fig.~\ref{cholesteric}b
. This sets the lower limit of the cholesteric pitch of Pf1 virus to the
size of the capillary. {\it The theoretical challenge is to explain
why two such similar chiral molecules can have extremely different
values of the cholesteric pitch}.

The concentrations of the isotropic to cholesteric phase transition
are quantitatively explained by the Onsager theory establishing
{\it fd} as an ideal model of hard rods~\cite{Tang95,Fraden95}. On
the other hand the origin and mechanism of the formation of the
cholesteric structure in liquid crystalline {\it fd} remains a
challenge. It is important to note that the Onsager theory predicts
equally well the concentration dependence of the isotropic-nematic
and isotropic-cholesteric phase transition. The reason for this
being that the free energy difference between the isotropic and
nematic phase is much larger than the free energy difference
between the nematic and cholesteric phase. The average twist in the
cholesteric phase between two neighboring molecules is generally
less then 0.1$^{\circ}$. This is negligible when compared to the
magnitude of director fluctuations of rod-like molecules in the
nematic phase, which are typically around 10 degrees.

In this paper we study in detail the continuum properties of the
cholesteric phase formed by {\it fd} virus, a charged semi-flexible
virus with contour length 880 nm, diameter 6.6 nm and persistence
length of 2200 nm. We measure the dependence of equilibrium pitch
on concentration and ionic strength and compare it to theory. For
two ionic strengths we also measure the value of the twist elastic
constant $(K_{22})$ as a function of concentration. Initial studies
of the cholesteric phase of {\it fd} can be found in the thesis of
Oldenbourg \cite{Oldenbourg81}

 \section{Experimental Results}

Bacteriophage {\it fd} was grown and purified using standard
techniques of molecular biology \cite{Maniatis89}. A stock solution
at a volume fraction of 5\% was dialyzed at room temperature
against Tris-HCl buffer at pH 8.1 of the desired ionic strength and
3 mM sodium azide $(\mbox{NaN}_{\scriptsize 3})$ was added to
prevent any bacterial growth. This solution was spun in an
ultracentrifuge which resulted in very viscous irridescent pellet
indicating a smectic or crystalline order. The pellet was
resuspended at 4$^{\circ}$ overnight in the amount of buffer so
that its final concentration was just above cholesteric-smectic
transition. A dilution series was made from the smectic to
isotropic phase. The concentration was measured using the
extinction coefficient $OD_{269 nm}^{1 cm}= 3.84$ \cite{Fraden95}.
Quartz x-ray capillaries of 0.7 mm diameter were cleaned with
sulfuric acid and repeatedly rinsed with de-ionized water before
being filled with {\it fd} samples. After 24-48 hours the sample
would equilibrate and a typical cholesteric texture was observed.
As described previously a {\it fd} suspension exhibits a increase
in the isotropic-cholesteric concentration over a period of few
weeks \cite{Tang93}. The origin of this time dependence is not
known (although we suspect bacterial growth). We also observed that
the cholesteric pitch systematically increases over the same time
period. Typically the values of the co-existence concentrations
shift by 1\% per week for first few months after the sample is
prepared~\cite{Tang96}. Because of this effect we performed
measurements on samples that were at most a few days old.

When viewed under a polarizing microscope the cholesteric phase
displays typical dark and bright stripes indicating that molecules
are perpendicular and parallel to the plane of polarizers,
respectively as is shown in Fig.~\ref{cholesteric}. The distance
between two bright lines is equivalent to half the value of the
pitch $P_0$. The microscope objective was focused on the mid-plane
of the cylindrical capillary and the picture was displayed on a
video monitor where the cholesteric pitch was measured using a
ruler. Measurements of pitch are very sensitive to defect and
boundary conditions and can vary by as much as 20\% within the same
sample. However, by repeating the measurement 15-25 times along the
full length of the sample it is possible to get a reproducible
value of the cholesteric pitch.

\begin{figure}
\centerline{\mbox{\epsfig{file=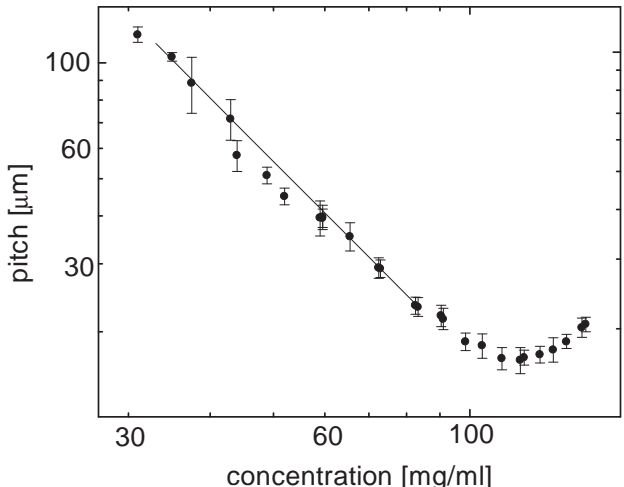,width=8.25cm}}}
\caption{\label{53mmconc} Log-Log plot of the cholesteric pitch $(P_0)$
versus concentration (c) of the {\it fd} virus.  The ionic strength
was 53 mM. Cholesteric pitch was measured from the fingerprint
texture which was observed in a sample contained in a cylindrical
capillary. Error bars were obtained by repeating the same
measurement along the full length of the capillary sample. A fit to
$P_0\propto c^{\nu}$ reveals the scaling exponent to be
$-1.65\pm0.05$. The data spans the entire range of concentrations
of the cholesteric phase for this ionic strength and temperature.
The sample is isotropic below 30 mg/ml and smectic above 150
mg/ml.}
\end{figure}

As the concentration of {\it fd} is increased we observe a sequence
of phase transitions from isotropic to cholesteric and to a smectic
phase \cite{Dogic97}. We measured the pitch of the cholesteric
phase for concentrations ranging from the isotropic to the smectic
phase. The typical behavior of cholesteric pitch as a function of
concentration is shown in Fig.~\ref{53mmconc}. We observe that the
pitch decreases with increasing concentration until it saturates at
a certain value. As the concentration of the virus increases
further and approaches the smectic transition we observe a slight
unwinding of the pitch. This unwinding of the cholesteric pitch
close to the smectic transition has been observed before in
thermotropic cholesteric liquid crystals \cite{Pindak74}. It is
generally assumed that it is due to pre-smectic fluctuations in the
cholesteric. Curiously, when we looked for these pre-smectic
fluctuations with light scattering we did not observe any
\cite{Dogic97}. At the cholesteric-smectic phase boundary the
cholesteric pitch abruptly unwinds. We repeated the measurements
for a  range of ionic strengths from 4 mM to 63 mM as shown in Fig.
\ref{pitchvsconc}.

By measuring the critical  magnetic field necessary to induce the
cholesteric to nematic phase transition it is possible to deduce
the value of the twist elastic constant
\cite{deGennes93,Meyer69,Meyer68}. The following formula (in c.g.s.
units) relates the critical field to the value of the twist elastic
constant $K_{22}$

\begin{equation} H_{c}=\frac{\pi}{2}
\left(
\frac{K_{22}}{\chi_{v}}
\right)^{\frac{1}{2}}q_{0}=\pi^{2}\left(\frac{K_{22}}{\chi_{v}}\right)^{\frac{1}{2}}\frac{1}{P_{0}}
\label{twistconstant}
\end{equation}

The value of the diamagnetic anisotropy $\chi_0 = \chi_\parallel -
\chi_\perp$ per molecule for {\it fd} is $7\times
10^{-25}$~$\mbox{erg}/\mbox{G}^2$ in c.g.s. units~\cite{Torbet81}.
To convert $\chi_0$ to the diamagnetic anisotropy per unit volume
$\chi_v$ used in Eq.~\ref{twistconstant} it is necessary to
multiply $\chi_0$ with the number of {\it fd} molecules per 1
$\mbox{cm}^3$ ($n$) given by $n = c N_{\mbox{\scriptsize A}} /
M_{\mbox{w}}$ where $c$ is the mass concentration of {\it fd},
$M_{\mbox{w}} = 1.64 \times 10^{7}$~g/M~\cite{Fraden95} is the
molecular weight of {\it fd}, and $N_{\mbox{\scriptsize A}}$ is
Avogrado's number. We note that the reported value of $\chi_0$ used
in this paper~\cite{Torbet81}, is an overestimate (by at most a
factor of two) of the actual value~\cite{ChiEstimate}.

\begin{figure}
\centerline{\mbox{\epsfig{file=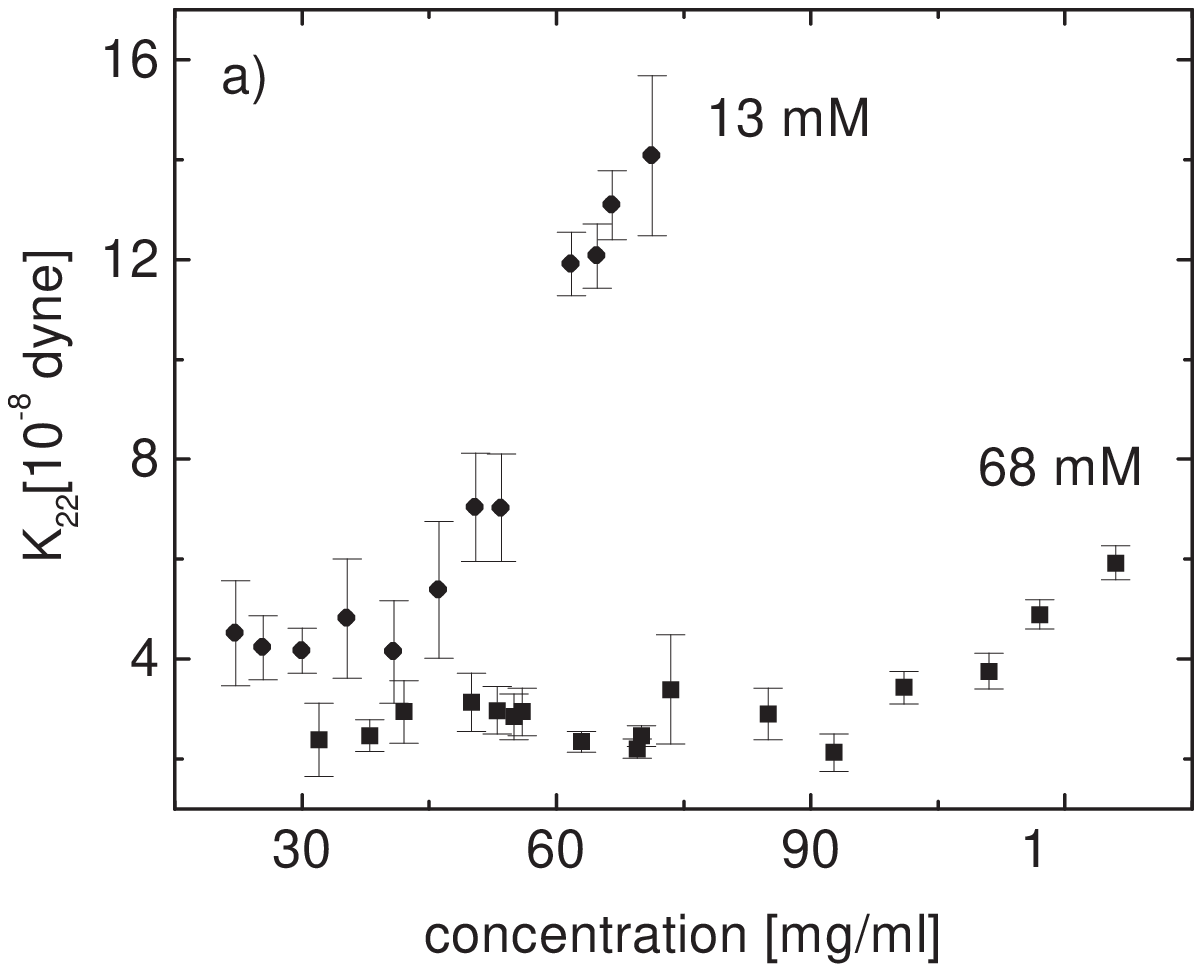,width=8.25cm}}}
\centerline{\mbox{\epsfig{file=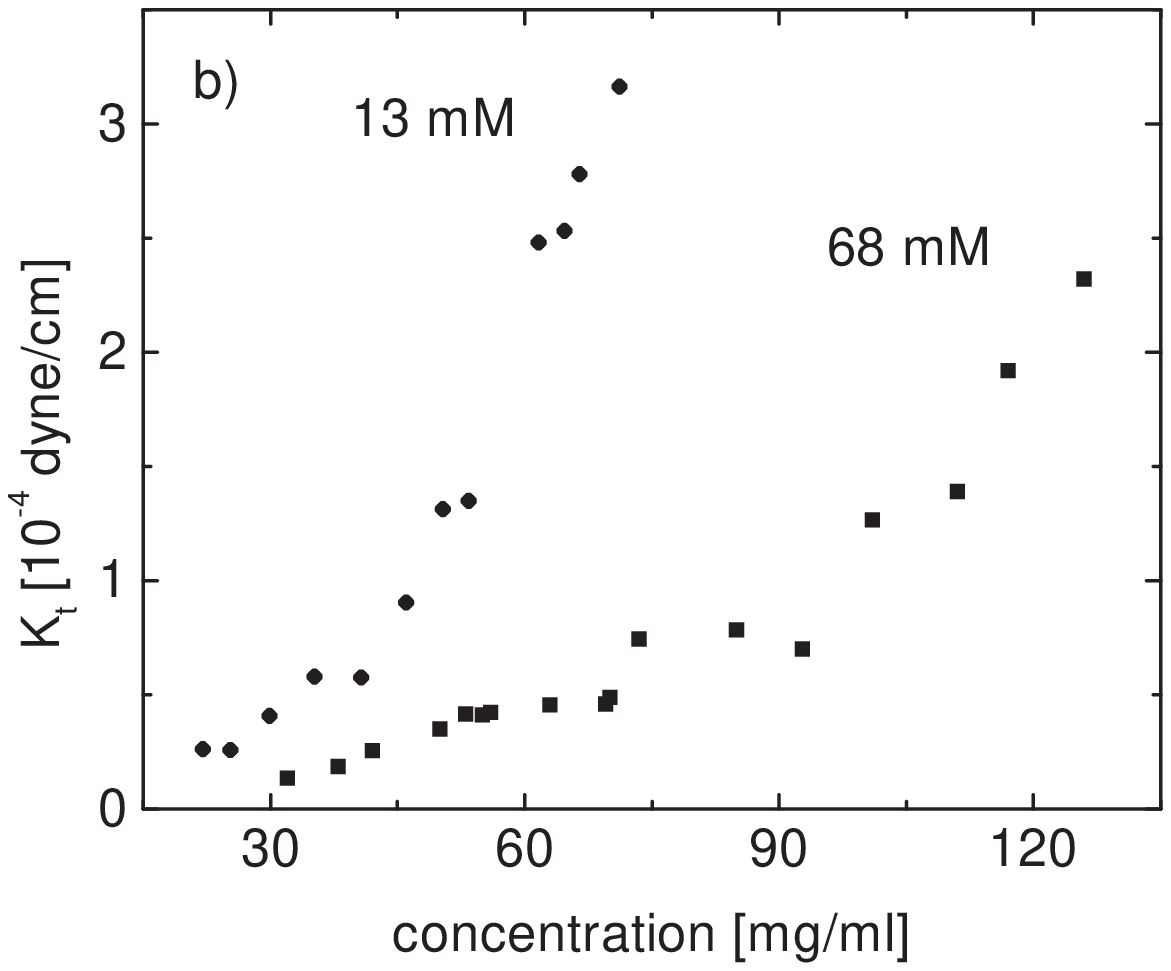,width=8.25cm}}}
\caption{\label{elasticconst} a)
Concentration dependence of the twist elastic constant $(K_{22})$
determined by measuring the critical magnetic field required for
inducing the cholesteric to nematic transition at two ionic
strengths of 13 mM and 68 mM. Error bars indicate a range between
lowest magnetic field required to unwind the sample and the highest
magnetic field at which there is still evidence of the cholesteric
phase. b) Chiral ``twist'' constant $(K_t)$ calculated from the
relation derived from continuum theory $K_{t}=\pi K_{22}/P_0$ for
two ionic strengths of 13 mM to 68 mM. Note that the sample
concentrations at higher ionic strength of 68 mM do not span the
entire cholesteric range. The reason for this is that the
accessible magnetic fields were not strong enough to completely
unwind samples with very high concentration
\protect\cite{Harris99}.}
\end{figure}

 We placed the sample with the long axis of the cylindrical
capillary parallel to the magnetic field and simultaneously
observed the unwinding of the cholesteric pitch under a microscope.
Theoretically we expect that the equilibrium pitch will scale as
the fourth power of applied field. Utilizing this sharp dependence
on magnetic field we measured the highest value of the field at
which the characteristic fingerprint cholesteric texture could
still be observed and the lowest field at which we observed no
cholesteric texture. Using Eq.~\ref{twistconstant} we can calculate
the range within which we expect the true value of twist elastic
constant. We have measured the value of the twist elastic constant
for various {\it fd} concentrations at two ionic strengths of 68 mM
and 13 mM as is shown in Fig.~\ref{elasticconst}.

\begin{figure}
\centerline{\mbox{\epsfig{file=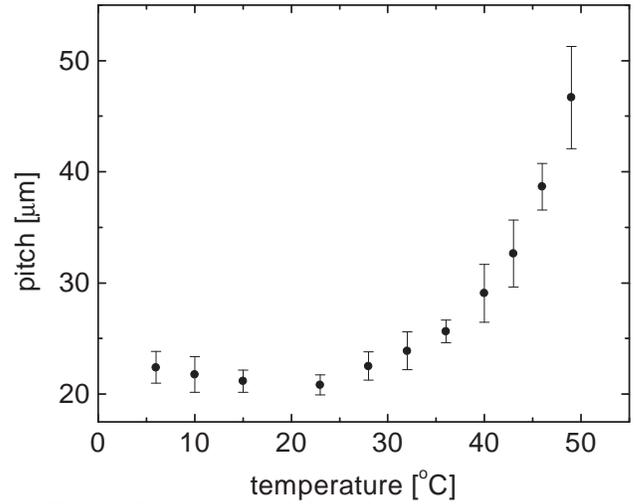,width=8.25cm}}}
\caption{\label{temppitch}
Dependence of {\it fd} cholesteric pitch on temperature. The rapid
increase in pitch at high temperatures is not due to denaturing of
the virus since the sample regains the original cholesteric
structure upon cooling to room temperature. The concentration of
the {\it fd} sample was 32.5 mg/ml and the ionic strength was 8mM.
The sample was far from the cholesteric-smectic phase boundary and
therefore the pitch unwinding cannot be due to pre-smectic
fluctuations. }
\end{figure}

The dependence of the cholesteric pitch on temperature is shown in
Fig. ~\ref{temppitch}. For this part of the experiment phosphate
buffer was used because of the smaller temperature dependence of
pK$_a$. As the temperature is increased we observe rapid unwinding
of the cholesteric pitch. The measurements are reversible and upon
a sudden temperature quench the sample winds up within a few
minutes as the cholesteric pitch attains its equilibrium value.

\section{Disscusion and Conclusion}

Bacteriophage {\it fd}, as most other biological colloids, has a
charged surface to maintain its stability in solution. Onsager
 was first to show that in the dilute limit the free energy of a
charged rod is approximately equal to a free energy of a neutral
rod with an effective diameter larger then its bare diameter
\cite{Onsager49,Stroobants86a}. It follows that the volume fraction
of the isotropic to nematic phase transition scales with
$D_{\mbox{\scriptsize eff}}$. The dependence of
$D_{\mbox{\scriptsize eff}}$ on ionic strength for {\it fd} was
calculated previously \cite{Tang95} and the concentration
dependence of the isotropic-cholesteric phase boundary on ionic
strength were previously investigated~\cite{Tang95}.

\begin{figure}
\centerline{\mbox{\epsfig{file=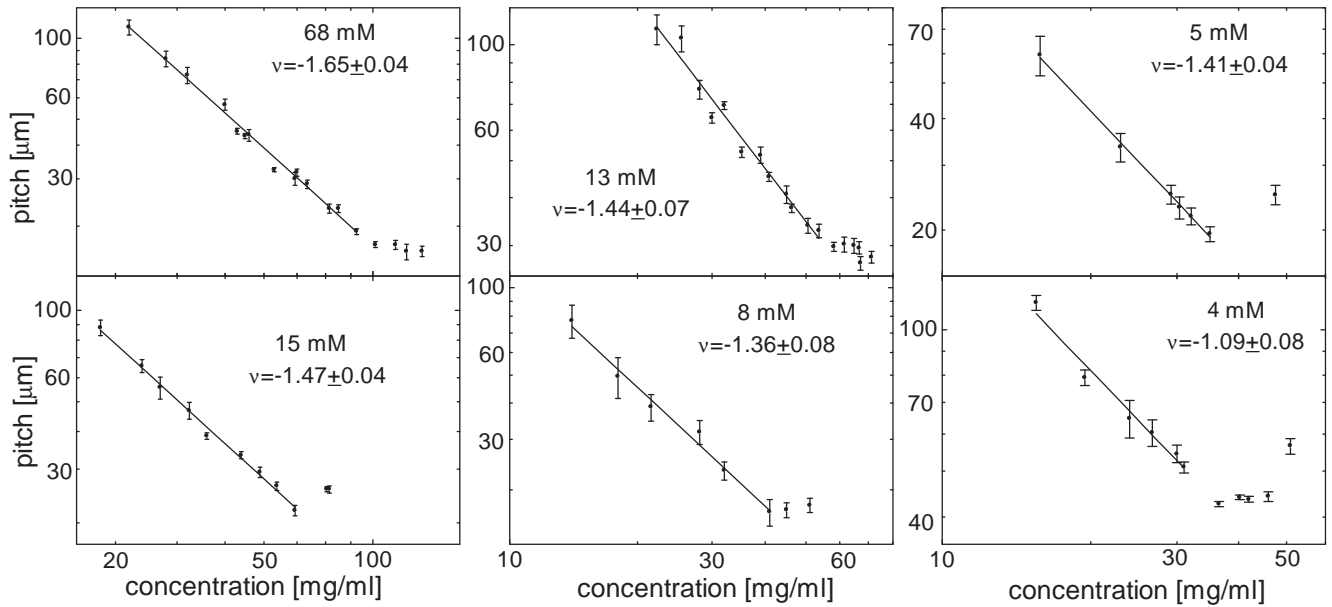,width=17.78cm}}}
\caption{\label{pitchvsconc} Log-Log plots of the cholesteric pitch
dependence $(P_0)$ versus {\it fd} concentration (c) for six
different ionic strengths ranging from 68 mM to 4 mM as indicated
in upper right corner of each graph. In each plot data was fitted
to an equation of form $P_0 \propto c^{\nu}$. The value of fitted
exponent ${\nu}$ is also indicated in upper right corner for all
six ionic strengths. At the highest ionic strength the exponent
$\nu$ equals -1.65$\pm0.04$ and systematically decreases with
decreasing ionic strength until it reaches the value of
-1.09$\pm$0.08 for the lowest ionic strength of 4mM. Each data set spans the
entire cholesteric phase from the isotropic boundary at low
concentrations to the smectic phase at high concentrations.}
\end{figure}

Motivated by the predictions from various theories
\cite{Straley76,Odijk87,Pelcovits96} we have tried to fit our
measurements of the pitch  to an exponential form $P_0 \propto
c^{-\nu}$ as shown in Fig.~\ref{pitchvsconc}, for ionic strengths
from 4 mM to 68 mM. At 68 mM ionic strength we find that the
scaling exponent $\nu$ has a value of $1.65\pm0.05$. Physically, at
high ionic strength the Debye screening length $\kappa^{-1}$
becomes very small and the effective diameter approaches the hard
rod limit \cite{Tang95}. However, even for 68 mM ionic strength,
$D_{\mbox{\scriptsize eff}}$ (13nm) is still considerably larger
then the bare diameter (6.6 nm). As the ionic strength decreases to
4mM, $D_{\mbox{\scriptsize eff}}$ increases to about 30 nm.
Experimentally, we find that the exponent $\nu$ systematically
decreases with decreasing ionic strength until it reaches the value
of 1.09$\pm$0.08 for low ionic strengths of 4mM.

Other experimental investigations of lyotropic cholesteric liquid
crystals involved the neutral synthetic polymer
poly-benzyl-L-glutamate (PBLG)~\cite{DuPre75,Uematsu84}, charged
DNA~\cite{VanWinkle90} and charged cellulose
suspensions~\cite{Dong97}. In the case of PBLG, the dependence of
pitch on concentration scales with exponent $\nu=-1.8$. In the
limit of high ionic strength we measure the value of the equivalent
exponent for {\it fd} to be
-1.65. It might be expected that as ionic strength is further
increased and {\it fd} increasingly becomes hard rod-like the
agreement between PBLG and {\it fd} data would be even better.
Experimental results for both {\it fd} and PBLG are close to the
theoretical prediction for the scaling constant $\nu$ due to Odijk
and Pelcovits which are
-1.66 and -2.0 respectively~\cite{Odijk87,Pelcovits96}.

For DNA and closely related synthetic double-stranded RNA there are
conflicting reports in the literature. Jizuka et. al.
\cite{Jizuka78}, in their study of double-stranded RNA, reported
the exponent to be -1.1. Senechal et. al. \cite{Senechal80} on the
other hand, finds that the scaling exponent for an equivalent
system to be -0.5. One possible explanation for this discrepancy is
the fact that the samples used had different ionic strength.
Senechal et. al. did their experiment in distilled water, while
Jizuka's experiments were done at an ionic strength of 100mM. This
agrees with our observations about the influence of ionic strength
on the scaling exponent. For DNA of contour length equal to the
persistence length, Van Winkle reported that the cholesteric pitch
is independent of the concentration \cite{VanWinkle90}. However, it
is also reported that DNA forms a pre-cholesteric phase with large
cholesteric pitch and undetermined structure.  This is in contrast
to the phase behavior of {\it fd} where there is no sign of any
transition between two cholesteric phases of different pitches.

Using a magnetic field induced cholesteric to nematic phase transition we have
measured the concentration dependance of the elastic constant $K_{22}$.
Although the data is noisy, the value of $K_{22}$ is almost constant in samples
of low concentration. Once the concentration of the virus is close to the
smectic phase concentration, the value of $K_{22}$ rapidly increases. It is
interesting to note that samples at low ionic strength have a significantly
greater plateau value of $K_{22}$ then the high ionic strength samples, in
contradiction to the theory of elastic constants that takes into account the
charge of rod-like particles~\cite{Vroege87}. Dupre~\cite{DuPre75} measured the
value of the twist elastic constant in PBLG and found that there is little
variation over the entire concentration range, which agrees with the plateau we
observe in {\it fd} samples. However, in PBLG there is no sudden increase in
$K_{22}$ at high concentrations. Measuring values of the twist elastic constant
$(K_{22})$ and the pitch for identical samples enabled us to obtain values of
the chiral twist constant $K_{t}$ using the relation $P_0=2\pi
K_{22}/K_{t}$~\cite{Harris99} as shown in Fig.~\ref{elasticconst}.

Sato~\cite{Sato93} and coworkers have found that the expression for the
cholesteric pitch splits into an entropic and enthalpic part. The entropic
part, due to hard core repulsive forces, is temperature independent while the
enthalpic part, due to attractive forces, scales inversly with temperature.
They measured the temperature dependence of the cholesteric pitch for a number
of liquid crystalline polymers and found that the pitch scales with inverse
temperature.  It is important to stress that their analysis is thermodynamic
and therefore disregards all complex details of molecular interactions.
Surprisingly, we find that the temperature dependence of the cholesteric pitch
of {\it fd} does not agree with this general thermodynamic formalism. A
possible source of this discrepancy lies in the assumption used by Sato et. al.
that changing temperature has no effect on the internal structure of the
polymer. While this is true for most synthetic polymers, it is well known not
to be the case for bioploymers. Viruses Pf1 and M13 undergo structural symmetry
transitions upon decreasing temperature below 8$^{\circ}$C and changing surface
charge, respectively \cite{Bhattacharjee92}. There are also indications that
the flexibility of {\it fd} depends non-monotonically on temperature
\cite{Tang96}. Any one of these factors could be the reason for the unusual
dependence of pitch on temperature.

\section{Acknowledgments}

This research is supported by National Science Foundation grant No.
DMR-9705336. Additional information is available online at
www.elsie.brandeis.edu

\widetext
\narrowtext

\end{document}